# Plasma and Thermal Processing Leading to Latitudinal and Temporal Variability of the Trapped $O_2$ at Europa and Ganymede

Apurva V. Oza,[1,2] Robert E. Johnson,[3,4] Carl A. Schmidt,[5] Wendy M. Calvin,[6] and Yuk L. Yung[1,2]

[1]Division of Geological and Planetary Sciences, California Institute of Technology, Pasadena, California, USA.
[2] Jet Propulsion Laboratory, California Institute of Technology, Pasadena, California, USA.
[3]Department of Astronomy, University of Virginia, Charlottesville, Virginia, USA.
[4]Department of Physics, New York University, New York, New York, USA.
[5]Center for Space Physics, Boston University, Boston, Massachusetts, USA.
[6]Geological Sciences, University of Nevada Reno, Reno, Nevada, USA.
E-mail: oza@caltech.edu , rej@virginia.edu

**Abstract**

We describe the physical processes affecting the formation, trapping and outgassing of $O_2$ at Europa and Ganymede. Following *Voyager* measurements of their ambient magnetospheric plasmas, laboratory data indicated that the observed ions mostly ejected from volcanic Io, would, in turn, impact and sputter their surfaces (Lanzerotti et al. 1978), decomposing the ice (Brown et al. 1982) and producing thin oxygen atmospheres (Johnson et al. 1982). More than a decade later Europa's $O_2$ atmosphere was inferred from observations of the O aurora (Hall et al. 1995) and 'condensed' $O_2$ bands were observed in Ganymede's icy surface at 5773 & 6275 Å (Spencer et al. 1995; Calvin et al. 1996). More than another decade later, their atmospheres were shown to have a dusk/dawn enhancement (Roth et al. 2016; Leblanc et al. 2017; Oza et al. 2019) confirmed by recent *Juno* data (Addison et al. 2024). Although the incident plasma produces these observables, processes occurring within the topmost surface are still not well understood. Here we note that the incident plasma particles produce a nonequilibrium defect density locally in the surface ice grains. Defect diffusion within these grains leads to the formation of voids and molecular products, some of which are volatile (Johnson and Quickenden 1997). Although some volatiles are released into the satellite atmospheres, others are trapped at defect sites or trapped in voids creating gas bubbles whose lifetimes (in steady state) are limited by the plasma-induced destruction rate. Here we discuss how trapping competes with annealing of the radiation damage. We describe the differences observed at Europa and Ganymede and roughly determine the observed trend with latitude of $O_2$ bands observed on Ganymede's trailing hemisphere (Trumbo et al. 2021). This understanding is used to discuss the relative importance of 'condensed' $O_2$ and $O_2$ adsorbed on regolith grains as atmospheric sources, accounting for dusk/dawn enhancements and temporal variability reported in 'condensed' $O_2$ band depths. Since plasma-induced damage and thermal annealing timescales drive oxidant variability on icy moons (likely also Callisto, Dione, and Rhea), they can help determine volatile downwelling, a potentially metabolic source for their oceans, and upwelling of other trapped oxidants (e.g. $CO_2$) suggestive of ongoing geologic activity.

## 1. Introduction

Energetic ions and electrons in the Jovian magnetosphere (Lanzerotti et al. 1978; 1983) impact the icy surfaces on Europa and Ganymede (e.g., Cooper et al. 2001; Nordheim et al. 2018) causing the sputter ejection of molecules from their surfaces (e.g., Johnson 1990; Teolis et al. 2017; Davis et al. 2021). The concomitant radiolysis is manifested by the production of thin atmospheres containing $H_2$ and $O_2$ (Johnson et al. 1982), products of the decomposition of ice in their surface grains (e.g., Reimann et al. 1984) as inferred from spacecraft and remote sensing observations (e.g., Hall et al. 1995; Roth et al. 2016; de Kleer et al. 2023). Equally interesting are the absorption bands seen in reflectance, also indicative of radiolytic decomposition of ice: 'condensed' $O_2$ (Spencer et al. 1995; Calvin et al 1996; Trumbo et al. 2021) seen also at Callisto (Spencer and Calvin 2002), $H_2O_2$ (e.g., Trumbo et al. 2019; 2023; Wu et al. 2024), $O_3$ (Noll et al. 1996; Ramachandran et al. 2024) and oxidized contaminants (Carlson et al. 2017) all trapped in the surface ice grains. The 'condensed' $O_2$ bands have been suggested to result from dimer absorption in solid $O_2$ (e.g., He et al. 2021; Michalsky et al. 1999), although some data suggests that the band shapes might be better fitted if the ice also contains $CO_2$ (Migliorini et al. 2022). These bands are also present in gas phase $O_2$, either by collision induced absorption (CIA) by pairs of $O_2$ or from van der Waals bound $(O_2)_2$ dimers. CIA requires densities far higher than what exists in the Galilean satellite atmospheres, and bound dimers in the gas phase are challenging to isolate in laboratory experiments. Since the satellite atmospheres are tenuous and the surface temperatures are such that $O_2$ is volatile down to the depth of the thermal wave, the 'condensed' $O_2$ bands were suggested to be due to $O_2$ trapped in radiation produced voids (Calvin et al. 1996; Johnson and Jesser 1997, hereafter JJ).

Indeed, since $O_3$ is not readily produced during the irradiation of ice, its presence in Ganymede's surface is consistent with its production by excitation of 'condensed' $O_2$ trapped in the large, near surface, ice grains (Johnson and Quickenden 1997, hereafter JQ). These observed bands are also consistent with the preferential loss of $H_2$ from the irradiated icy surfaces as suggested by the ambient plasma measurements (e.g., Cooper et al. 2001; Szalay et al. 2024) and modeling (e.g., Carberry Mogan et al. 2022). Below we refer to energetic particles generically as a plasma, acknowledging the term generally refers to bulk electrons and ions in the satellite tori.

Since *individual* $O_2$ molecules trapped at defect sites or in vacancies and voids are difficult to detect, the 'condensed' $O_2$ bands, which are associated with the excitation of pairs of $O_2$ molecules have proven to be useful. However, these bands, associated with multiple $O_2$ inclusions and often referred to as dimer excitations, do not account for all the $O_2$. Therefore, in addition to $O_2$ being driven into the satellite atmospheres and $O_2$ trapped *in* ice grains, we suggest *that* $O_2$ is transiently adsorbed and desorbed on grain surfaces and exists as a gas permeating their porous regolith (Johnson et al. 2003). That is, the spatial and temporal variations in the ambient atmospheres have been suggested to be due to *thermal release* of individual $O_2$ trapped at defects (e.g., Johnson et al. 2019; Teolis and Waite 2016). Interestingly, the 'condensed' $O_2$ band depths exhibit significant temporal variability with times short compared to the satellite orbital periods (e.g., Spencer et al. 2019; Trumbo et al. 2021). This aspect confirms that radiation processing is active, producing both gas-phase and trapped species in competition with their destruction and thermal annealing of the radiation damage.

Motivated by such observations, and by the modeling of the dusk/dawn atmospheric enhancements (Oza et al. 2018; 2019), confirmed by recent magnetic field observations by *Juno* (Addison et al. 2024), below we discuss the physics of $O_2$ trapping in and its ejection into the very thin atmospheres on Europa and Ganymede, while addressing the observed temporal and spatial variations. We first review aspects of the now extensive laboratory data and then describe our understanding of 'condensed' $O_2$ formation as well as its relevance to the satellite observations. Useful satellite data is given in Table 1 and the activation energies of the relevant physical processes are provided in Table 2.

**2. Radiolysis**

Whereas crystalline ice has an equilibrium density of defects determined by the temperature (e.g., de Koning 2020), incident energetic ions and electrons produce an over-density of displacements and interstitials forming a 'track' of vacancies and radicals along their paths through ice. The region around the 'track' is transiently heated by the excess energy deposited leading to diffusion of these defects and reactions. After that transient heat dissipates, defect diffusion continues, determined by the background temperature. As the solid anneals, the vacancies and interstitials can recombine (annihilate) or are incorporated into various sinks. The diffusing radicals reacting with vacancies contribute to annealing (i.e., recrystallization) or they can react to form new chemical species. The diffusing vacancies can also accumulate producing voids (e.g., Okada et al. 2020). Voids produced by electron irradiation are readily seen in electron microscope studies of ice up to the peak temperatures on Ganymede, with their production and growth affected by impurities (Heide & Zeitler 1985). During annealing small voids can be absorbed by larger voids reducing the net surface to volume ratio, a process often referred to as Oswald ripening. This process also produces segregated precipitates in solids (e.g., Gusak et al. 2006). Void disappearance with increasing temperature is seen in studies of small grain, microporous ice, presumably due to mobility in the presence of extensive grain surfaces. Diffusion and reactions are, of course, enhanced along the pathway formed by penetrating radiation (Benit and Brown 1990) as well as along surfaces: e.g., the atmosphere interface, grain boundaries, and internal surfaces in voids. Such processes have been extensively discussed, particularly for room temperature solids exposed to energetic particle radiations (e.g., Wiedersich 1972; Jiang et al. 2022) and the early studies of the radiolysis of ice are summarized in JQ.

TABLE 1. RELEVANT SATELLITE DATA FOR EUROPA AND GANYMEDE'S SUNLIT TRAILING AND LEADING HEMISPHERES

| Region | $T_{eq}$ [a] (K) | $\overline{T}$ [b] (K) | Thermal Annealing Time/Orbital Period [c] | Plasma Flux , $f_{ion}$ [d] (ions/s/ cm$^2$) | Plasma Bubble Impact time/ Orbital Period [e] | $\overline{A}_b$ Bond Albedo [f] | $O_2$ Band Depths (%) [g] |
|---|---|---|---|---|---|---|---|
| E (T) | 135 | 126 | 0.06 | $3 \times 10^8$ | 0.02 | 0.45 | ~0.3 |
| E (L) | 125 | 116 | 1.4 | $10^8$ | 0.06 | 0.65 | ~0.3 |
| G (T) | 147 | 137 | $1.2\times10^{-3}$ | $6 \times 10^4$ | 53 | 0.37 | ~2 |
| G (L) | 142 | 132 | $4.2\times10^{-3}$ | $10^7$ | 0.3 | 0.43 | ~0.7 |

[a] Spencer 1987 (equatorial values)

[b] Average hemispheric temperatures: $\overline{T} = \int_0^{\frac{\pi}{2}} T(\theta) cos(\theta) d\theta$

[c] Cao 2021 (Schmitt et al. 1989 Fig. 12). Also Fig. 2 modeled as $t_{anneal} \sim 10^{13}$s $e^{(-0.46 eV/kT)}$

[d] $f_{ion}$: Plasma ion flux (Poppe et al. 2018)

[e] Plasma ion impact time for a 4nm bubble [$t_b \sim (\pi r_b^2 f_{ion})^{-1};$]. Proportional to the average bubble destruction time (Appendix) within the average plasma penetration depth.

[f] Spencer 1987

[g] Spencer et al. 1995; Calvin et al. 1996; Spencer and Calvin 2002; Trumbo et al. 2021 (rough averages)

TABLE 2. ACTIVATION ENERGIES DISCUSSED: EXP($-E_A/kT$)

| Process | $E_a$(eV) |
|---|---|
| $E_{ads}$ $O_2$ adsorption on grain surfaces[a] | ~0.11-0.14 |
| $E_{vm}$, vacancy formation in ice[b] | ~0.34±0.07 |
| $E_{plasma}$ Plasma particle ejection of $O_2$[c] | ~0.03-0.06 |
| $E_{anneal}$ Crystallization rate: lab samples[d] | ~0.46 |
| $E_{O2}$, 'Condensed' $O_2$ Ganymede[e] | ~0.3 |

[a] Johnson et al. (2019) Europa Dusk/dawn enhancement
Teolis and Waite (2016) Dione, Rhea seasonal variability
Lab data: Laufer et al. (2017); He et al. (2016)
[b] Experimental values in Eldrup (1976)
[c] Reimann et al. (1984), Teolis et al. (2017):
from fits of ice sputtering yield of $O_2$ vs. temperature
[d] Schmitt et al. (1989); Cao (2021)
[e] This article based on Trumbo et al. (2021) equatorial data

The diffusion of H atoms, produced by proton implantation or by molecular dissociation, has been shown to form $H_2$ in several materials (e.g., Behrisch and Eckstein 2007) including ice (Christianson and Garrod 2021). Depending on the temperature, the resulting $H_2$ can also diffuse efficiently (e.g., Patterson and Saltzman 2021), often along grain boundaries, and can desorb at an external surface. However, when $H_2$ is formed in or enters a void, it can become more stably trapped forming a gas bubble. This is well studied and leads to the swelling of highly irradiated, room temperature solids (e.g., Vook et al. 1975) as well as to the formation of $H_2$ bubbles in UV irradiated low temperature ice (Tachibana et al. 2017). Therefore, data on fast proton and alpha particle production of $H_2$ and He gas bubbles in reactor walls was used early on (e.g., JJ) to help understand the 'condensed' $O_2$ observations at Ganymede. That is, like H, radiolytically produced O can diffuse along surfaces or defect pathways and react at defect sites producing transiently trapped $O_2$. Or, if the reaction occurs on a surface in a void, it can lead to more stably trapped $O_2$, which we will refer to at times as a bubble. The presence of oxygen bubbles in Antarctic and Arctic ices, formed by co-deposition, is well established (e.g., Miller 1969), and even confirmed in thin samples (e.g., Bahr et al. 2001). However, interior $O_2$ bubbles have not been directly identified in radiation processed ice. Such formation and trapping of $O_2$ was suggested early on by its release on warming relatively thick irradiated samples (e.g., JQ). Also a luminescence feature in photolyzed ice was used to suggest that $O_2$ can form at an interior surface from diffusing O atoms (Matich et al. 1993). But convincing laboratory studies are still needed. Although attempts have been made to explain the formation of $O_2$ in irradiated ice using gas phase reaction rates (e.g., Li et al. 2022), the relevant processes in solids differ significantly. That is, even at the highest temperatures on Ganymede (~160K), $O_2$ is *not* soluble in crystalline ice so that defect sites and surfaces play a critical role (e.g., appendix of Johnson et al. (2003). Therefore, below we focus on processes occurring in solids.

Because of the relatively efficient diffusion of H in ice at satellite temperatures, its subsequent desorption from irradiated samples enhances the production of oxidants (Brown et al. 1982). This has been confirmed by many subsequent experiments, as summarized in Teolis et al. (2017), and rediscovered regularly (e.g., Abellan et al. 2023). Of course, in the presence of carbon, sulfur, etc., other oxidized species are produced due to the preferential loss of $H_2$, as seen in reflectance spectra (e.g., Carlson et al. 2017). Laboratory studies, indicate that the production and ejection of $O_2$ exhibits a dose dependence consistent with the dissociation, on average, of two water molecules per $O_2$ ejected (e.g., Reimann et al. 1984) *or* with the formation and the excitation of a precursor (Sieger et al. 1998; Orlando and Sieger 2003). In addition, the $O_2$ yield is small and independent of T below ~60-80K (e.g., Teolis et al. 2017; Davis et al. 2021), roughly consistent with defect mobility becoming efficient above ~ 0.3– 0.5 times the melting temperature (e.g., JJ). With increasing temperature, the ejection of $O_2$ from irradiated laboratory ice samples increases, indicative of an activated process, but with an unexplained activation energy (e.g., Reimann et al. 1984). Here we suggest that the extracted activation energy is determined by the surface tension of ice (see Appendix A1).

Primary products are formed in irradiated ice other than those discussed above. As extensively observed, the OH produced can react with impurities or react to produce $H_2O_2$ (e.g., Mifsud et al. 2022) a species seen in irradiated ice samples as well as in the reflectance spectra of Europa and Ganymede (e.g., Trumbo et al 2021). Although not volatile, when formed in voids, $H_2O_2$ inclusions can also build up. Therefore, Cooper et al. (2003) suggested that $H_2O_2$ molecules trapped in voids might react with increasing temperature, or be excited by radiation, contributing to the production of $O_2$ bubbles, a suggestion not yet confirmed. Although $O_2$ formation and ejection increases with increasing ice temperature (e.g., Reimann et al. 1984), above ~150-160K rapid recrystallization occurs and the $O_2$ yield drops. In contrast, $H_2O_2$ appears to be a stable product at lower temperatures (e.g., Teolis et al. 2017). Indeed, at Ganymede, peroxide is preferentially concentrated in the cold polar regions (e.g., Bockelée-Morvan et al. 2024), while the 'condensed' $O_2$ bands are seen to be dominant at the warmer lower latitudes. At Europa, however, peroxide is also seen in some low latitude regions, possibly collocated with other trace species such as $CO_2$ (Trumbo et al. 2019; Wu et al. 2024). Therefore, the solid-state chemistry occurring in ice (Johnson et al. 2003) still requires considerable work. Ice irradiation experiments at ~10 keV suggest peroxide-rich regions on Europa (e.g. Tara and Powys regions) are radiolytically enhanced by the presence of $CO_2$ (Mamo et al. 2025) likely affected also by the available oxidant trapping described here. After we describe below how irradiation-induced $O_2$ can be formed and trapped in voids, using data in Tables 1 and 2, we consider its contribution to the icy satellite atmospheres.

### 3. $O_2$ Bubble Formation: Satellite vs. Laboratory Surfaces

Although laboratory studies show that the radiation-induced decomposition of ice leads to the formation of $O_2$ and $H_2$, the absence of direct detection of the 'condensed' $O_2$ bands in irradiated samples remains a problem. The samples typically studied differ significantly in porosity and grain size from those on satellite (e.g., Cassidy and Johnson 2005) and comet surfaces (Oza and Johnson 2024). The vapor deposited laboratory samples tend to be small grain, microporous and thin, in order to reduce charging by the incident particles. On the other hand, the reflectance data for the icy Galilean satellites suggest their surfaces are composed of large, primarily crystalline, grains of ~ 100μm to 1mm (e.g., Ligier et al. 2019; Stephan et al. 2020; King and Fletcher 2022). These surfaces are presumed to be fairy castle structures with a very high pore space between such grains. Because of their high porosity, the satellite regoliths are permeated with atmospheric volatiles and the area accessible to incident radiation is much larger than the satellites' geometric surface.

Surface temperatures and 'condensed' $O_2$ band depths are extracted from the reflectance at visible wavelengths. However, these spectra depend on grain size, porosity, contaminant concentration and radiation damage, so that interpreting the reflectance data, though critical, is not simple. For instance, since crystalline ice is transparent in the visible, surfaces observed in reflectance can appear dark or bright in the visible for a number of reasons: dark (e.g., significant concentrations of absorbing contaminants or highly transparent with few absorption and scattering sites) and bright (e.g., small grains or large grains with a high density of defects acting as scattering sites). Surfaces that are bright in the visible can also be produced by radiation damage (e.g., Johnson 1985; Fama et al. 2010) or by returning $H_2O$ ejected by sublimation, sputtering or venting (e.g., Li et al. 2020) producing a fine grain frost (Trumbo et al. 2023; Villanueva et al. 2023). Although modeling indicates that $O_2$ can be formed and trapped, even at very low temperatures in small grains in irradiated ices in the interstellar medium (Jin and Garrod, 2020), the satellite surfaces differ considerably. Therefore, in the following we assume the observed 'condensed' $O_2$ is primarily formed and trapped in voids produced by irradiation of relatively large ice grains. We imagine a hierarchy of sizes: average grain size >> visible wavelength >> average bubble radius. We also presume that plasma particles producing the 'condensed' $O_2$ are primarily those with penetration depths comparable to the depths sampled by the reflected photons.

In addition to the physical differences discussed, radiation *dose rates* in laboratory studies are many orders of magnitude larger than those experienced by grains in the icy regoliths. Whereas *thermally* driven recrystallization rates in laboratory studies of irradiated low temperature ice samples are typically not relevant, that is not the case for all regions of the icy Galilean satellite surfaces. Because the fate of radiation damage is determined by a variety of diffusion processes, here we use the thermal annealing rate as a crude proxy for the processes occurring following radiation damage of ice (Table 3). Although crystallization studies of amorphous or vapor deposited ice exhibit significant scatter (e.g., Cao 2021; Baragiola 2003), we use the result in Schmitt et al., (1989) as a guide, which is roughly consistent with an activation energy ~0.46eV. These data suggest a large range of annealing rates across the satellite surfaces, varying from minutes at the peak temperature on Ganymede's trailing hemisphere to on the order of the orbital period at Europa and at high latitudes on Ganymede, as indicated in Table 1. Therefore, the lack of a direct laboratory detection of the 'condensed' $O_2$ bands in irradiated ice samples we suggest is due to the disparity between annealing and radiation damage rates, as well as the grain size issue discussed above. Although critically important, the application of the available laboratory data to the icy satellite observations is *not* straightforward.

Early studies of the release of $H_2$ and $O_2$ on warming of thick samples exposed to penetrating radiation are suggestive (JQ), new studies, possibly using electron microscopy (e.g., Abellan et al. 2023), on relatively thick annealed samples are needed. This is critical as the absence of direct laboratory observations of radiolytically induced 'condensed' $O_2$ bands has led to the suggestion that $O_2$ cold traps exist in the topmost ice layer of the icy Galilean satellites (e.g., Vidal et al. 1997; Baragiola and Bahr 1998). However, even for porous icy regoliths, the thermal skin depths are several cm, implying that low temperatures sufficient to form $O_2$ are unlikely. This is especially the case at low latitudes on Ganymede's sunlit trailing hemisphere where 'condensed' $O_2$ band depths are strongest at the equator ~147 K.

Since incident plasma particles transiently supersaturate the local defect density, the growth of voids resulting in bubbles of trapped volatiles is governed by the flux of vacancies vs. the recrystallization rate (e.g., JJ; Jiang et al. 2022) as discussed above. However, incident radiation penetrating a gas-filled void (bubble) can also cause *dissolution of the bubble and dispersion of the trapped species*. That is, bubble damage, which is typically ignored, competes with aggregation and growth. Under steady irradiation, a quasi-equilibrium is established: higher dose rates produce a high density of smaller, on average, bubbles that have higher interior surface energies (e.g., Gittus 1978; Tachibana et al. 2017). Therefore, in the presence of annealing, the *average* bubble dimension varies inversely, but slowly, with the radiation damage rate. Noting that the defect production rate is roughly proportional to the energy deposition rate, dE/dt, JJ referred to an estimate of the *averag*e radius, $r_b$, of gas *'bubbles'* produced under steady energetic particle implantation as

$$r_b \alpha \ [D(T)/(dE/dt)]^x \quad (1)$$

Here x is a small exponent (x~1/4 in Gittus (1978)) and D(T) accounts for the relevant diffusion processes. In our discussions below, the temperature dependence of the recrystallization rate of ice is used as a proxy for these processes. In addition, a rough estimate for the average *void* radius under steady irradiation, accounting for growth by vacancy diffusion *and* void destruction, is derived in Appendix A2. It has a similar form to that in Eq. 1 but with x~1/2, verifying the inverse dependence on the irradiation rate. These expressions suggest that a low flux of highly penetrating plasma particles, with a lower energy deposition per unit path length, penetrating a higher temperature ice can lead, on average, to larger bubbles.

Whereas Europa's surface experiences a more intense and diverse plasma flux, at low latitudes, Ganymede's field tends to deflect the heavy thermal ions that cause surficial damage and brightening. In the presence of higher average surface temperatures and lower fluxes of primarily penetrating particles, larger band depths on Ganymede should not be surprising. These 'low latitude' observations, discussed below, appear to include the open-closed field line boundary region (Fig.1 in Trumbo et al. 2021) in which the *Juno*-JADE instrument detected a flux of 30-100 keV electrons (Ebert et al. 2022). From *Galileo* encounters in the Jovian plasma sheet, simulations also suggest Ganymede's mini-magnetosphere shields electrons below 40 MeV at the equator (Liuzzo et al. 2020), which, based on Eq.1, enhances bubble formation. In addition to producing the observed O-aurora, these electrons, 4.5 keV-100 MeV, primarily deposit their energy in the surface ice (Waite et al. 2024) to depths (~10-100μm) roughly consistent with the grains being modified by *lightly* ionizing radiation.

Calvin et al. (1996) compared the visible bands observed on Ganymede's trailing hemisphere to those of 'condensed' $O_2$. Based on the band shape and the local temperature, they suggested the best fits were either solid-γ or liquid $O_2$, and estimated absorbance maxima from the data in Landau et al. (1962: $\alpha_D \sim 1.15$ or 4.8/cm respectively). Estimating the photon path length from reflectance data, they obtained ~0.1-1% concentration of $O_2$ to $H_2O$. Subsequently, Hand et al. (2006) used the depth, $\delta I$, of the 'condensed' $O_2$ band and the intensity, $I$, of the visible reflectance to estimate the fraction of $O_2$ in Europa's surface ice grains. Assuming wavelengths small compared to the grain radius, $r_g$, but large compared to internal scattering sites, they wrote

$$\delta I/I \sim (\alpha_D f_{O2}^2)\,[(4r_g/3)\,(1+r_0)^2/(1-r_0)] \quad (2)$$

The term in square brackets is an estimate of the average path length of the reflected visible photons determined by the average grain radius, $r_g$, and by internal scattering centers indicated by a 'reflectance' parameter $r_0$. Although they referred to a clathrate, they used an estimate for $\alpha_D \sim 1.6$/cm based on solid-γ $O_2$ at 0.577μm with $r_0 \sim 0.97$, obtaining $f_{O2}$ ~1.2-4.7% in Europa's surface ice (Hand et al. 2006; Hesse et al. 2022). If correct, then 'condensed' $O_2$ is clearly the dominant reservoir of $O_2$ in the near surface region (e.g., Johnson et al 2003). But $(\alpha_D f_{O2}^2)$ in Eq. 2 suggests individual $O_2$ molecules are randomly distributed within the grains, leading to an overestimate of condensed $O_2$ fraction. Therefore, $f_{O2}^2$ should be replaced by $f_{\gamma O2}$, the fraction of the pathlength in solid-γ $O_2$ which would give $f_{\gamma O2} < \sim 0.2\%$ at Europa. However, the applied parameters have significant uncertainties and the brighter surface at Europa has a sampling depth in the visible that differs from that at low latitudes on Ganymede. Assuming similar grain sizes and absorption coefficients, the ratio of their observed average band depths (~6.7) times the ratio of their albedos (0.67) suggests, very roughly, a 'condensed' $O_2$ fraction on Ganymede's trailing hemisphere in the region sampled by the reflected photons is ~4.5 times higher than the fraction in the region sampled by the reflected photons at Europa.

Since the pathlength through the regolith grains in the visible is critical, more detailed modeling is needed using new absorption coefficients (Calvin 2025). This is especially so since the related infrared bands (e.g., Bockelée-Morvan et al. 2024) have not been detected as discussed in Calvin et al. (1996). Recently Tinner et al. (2024) used steady state sputtering measurements to estimate a broad range of possible $O_2$ concentrations in irradiated samples for a range of grain sizes, but did not obtain dimer concentrations.

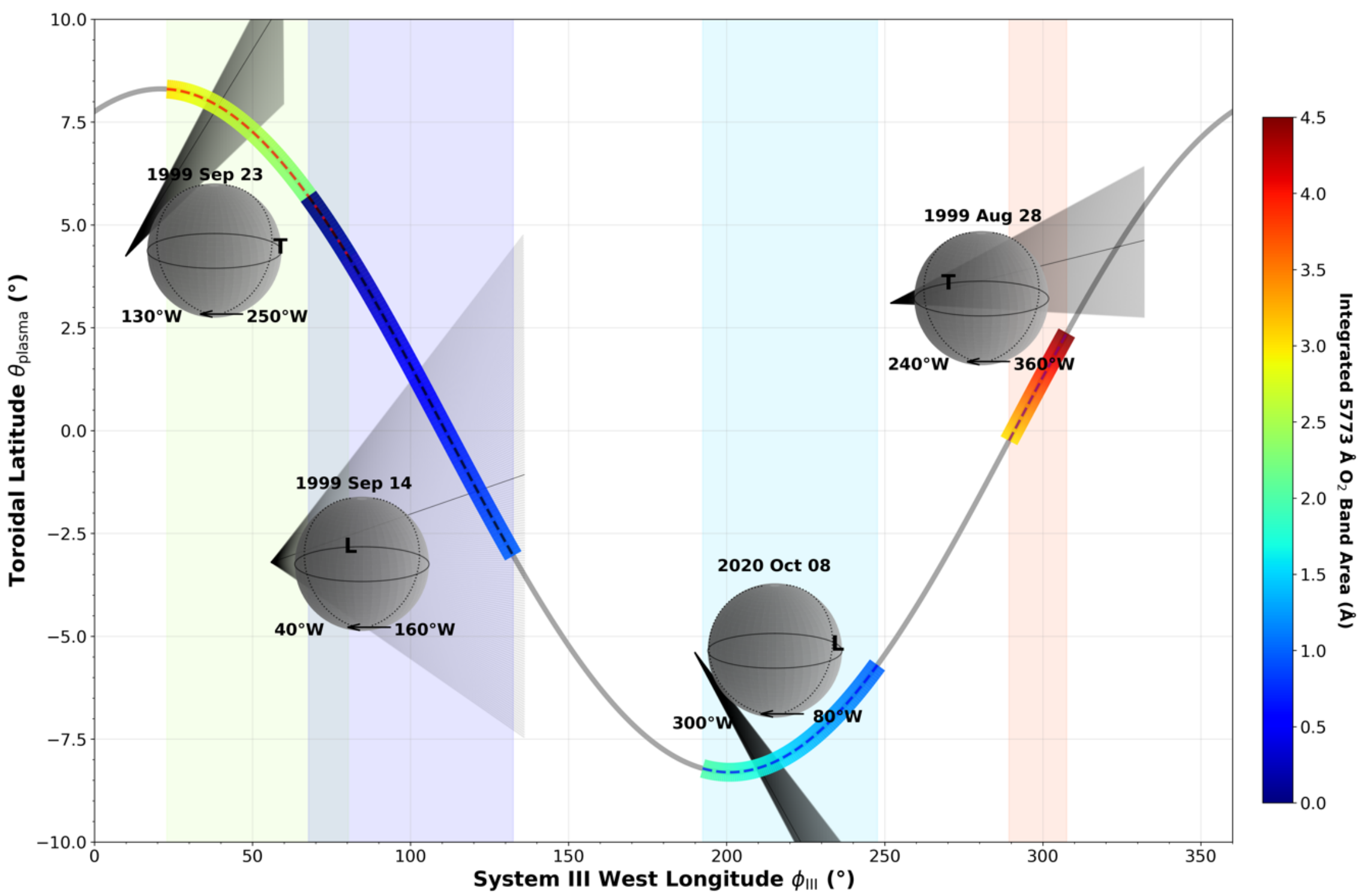

**FIG. 1.** Jovian plasma sheet latitudes $\theta_{Plasma-Torus}$ vs $\phi_{III}$ at Ganymede (dashed) during 4 observing periods spanning 1999-2020 summarized in Trumbo et al. (2021) suggestive of differences in energetic particles fluxes. The date of each observation, probed Ganymede W longitudes $(W)$ and approximate integrated $O_2$ band area $dA$ (centered at 5773 Å) are indicated in the colorbar, maximum at Ganymede's trailing (T) 1999 Aug 28 ( $\geq$ **4** Å) decreasing near Ganymede's leading (L) 1-2 Å (2020), 0.5 Å (1999), and back to 2.5 Å (1999 Sep 23). All 1999 observations are near Ganymede's equator at latitudes $\boldsymbol{\theta_G(1999) \sim 3.2}$°N, and $\boldsymbol{\theta_G(2020) \sim -1.4}$°S. The approximate torus angle is indicated locally at Ganymede for visualization.

**4. Ganymede Low Latitude Observations**

Focusing on the available observations, we note that *voids* are potential sinks for trapping diffusing radicals, which can lead to the formation of more stably trapped $O_2$, as discussed above. Void growth, controlled by vacancy migration, is an activated process depending on the background temperature (e.g., Eldrup 1976). In their summary of 'condensed' $O_2$ observations at Ganymede, Trumbo et al. (2021; Fig. 3A) noted a very rough correlation of a *drop* in the 'condensed' $O_2$ band area (~ factor of 3) at low latitudes with an *increase* in the albedo (~0.39→0.47). This trend is consistent with the darker trailing hemisphere having larger band areas on average than the brighter leading hemisphere (Calvin et al. 1996), as well as at the brighter high latitudes. These observations suggest that the regolith temperature plays a critical role affecting 'condensed' $O_2$ band formation as discussed above. That is, at low latitudes on Ganymede, where the annealing rate is rapid compared to its rotation rate (Table 1), the variation of the albedo is consistent with a small *average* temperature drop, $\Delta T(\phi)$, over the observing region centered on longitude $\phi$ from that centered on regions near $\phi \sim 270^{o}$. If the local temperatures are roughly in radiative equilibrium, then at longitude $\phi$, $dT(\phi)/T(270^{o}) \sim (1/4)\, dA)/[1-A(270^{o})]$ in which A is the visible albedo. Using the values above, the change in albedo from near $270^{o}$ to near $360^{o}$ West longitude is consistent with a fractional temperature change of ~5K, $\Delta T(360^{o})/T(270^{o}) \sim 0.037$, close to the average temperature change suggested by observations (Table 1). Assuming the 'condensed' $O_2$ bands are determined in part by thermal processing with an activation energy $E_{O2}$, and initially ignoring differences in the plasma flux and photon sampling depth, we approximate an averaged e ratio of the observed band areas as,

$$< R(\phi, 270 > \sim e^{-\left(\frac{E_{O_2}}{kT(270°W)}\frac{\Delta T(\phi)}{T(270°W)}\right)} \qquad (3)$$

Using the temperature ratio above suggests an effective activation energy $E_{O2} \sim 0.3$eV. This is a value close to the experimental value for activation energy of vacancy migration ($E_{vm} \sim 0.34 \pm 0.07$eV; Eldrup 1976), that determines the growth rate of voids in which, we suggest, 'condensed' $O_2$ is formed and trapped.

Of course, the particle flux, background temperature and photo-sampling depth all factor into the 'condensed' $O_2$ formation process. Therefore, it might not be surprising that in a different Ganymede observing period a *similar fractional drop in band area* was observed in going from near $270^{o}$ to $180^{o}$ West longitude, but the band areas were somewhat smaller (Trumbo et al. 2021; Fig. 3B). Since the temperature excursion is close to that estimated above, we suggest that drop in band area is due to a difference in the plasma flux. That is, although the background temperature can affect the net 'steady state' density of the gas-containing voids, the bubble formation *and* destruction rates depend on the radiation damage rate compared to the annealing rate (Appendix A2). Plasma variability at Ganymede is primarily due to the tilt of Jupiter's dipole and Io's activity. Ignoring the latter, Fig. 1 shows the plasma torus latitudes at Ganymede during the observing periods summarized in Trumbo et al. (2021). Although a simple correlation with magnetic latitude was *not* observed for Europa's brighter surface with weaker band depths (Spencer et al. 2019), Ganymede's optical aurorae roughly follow the expected fall off in electron density with latitude during eclipse $\phi_{obs} \sim 0^{o}$ (Milby et al. 2024), which implies a difference in the radiation rate of the surface ice grains (Waite et al. 2024). The optical aurorae show variability similar to *in-situ* data of the Io plasma torus at $\phi_{III} \sim 115^{o}$and $295^{o}$ (Bagenal et al. 2015) when the plasma sheet is ~ 0.4 – 0.7 $R_J$ ($\theta_{Plasma\ Torus} \sim 2$-$4^{o}$). Although suggestive of electron density and magnetospheric variations, a better correlation is needed and a more detailed description is required. This includes accounting for the shielding by Ganymede's magnetic field, as well as incorporating electron number density oscillations as measured by *in-situ* spacecraft such as *Voyager, Cassini* and *Galileo* (Bagenal et al. 2015). Ultimately, a better understanding at the molecular level of the radiation damage, void growth and annealing rates within the photon sampling depth is also needed.

Finally, it has been shown that the gas pressure in a bubble, $p_b$, can affect its stability (e.g., JJ). Under steady irradiation $p_b$ can build in a bubble of radius $r_b$ until it exceeds the interior surface tension, $\gamma_s$. For much higher pressures, the gas molecules would tend to become soluble, so that

$$p_b < \sim 2\, \gamma_s / r_b \qquad . (4)$$

Using $\gamma_s \sim 80 mJ/m^2$ and $p_b \sim n_{O2} kT$, with an average temperature on Ganymede's trailing hemisphere (~ 147K), the maximum column of $O_2$ confined per bubble is $\sim(r_b\, n_{O2}) < \sim 10^{16} O_2/cm^2$. If we assume that $n_{O2}$ approaches the liquid density, then $r_b < \sim 0.04 \mu m$. Of course, as is well documented in the Earth's atmosphere (McKellar et al. 1972), the 'dimer' absorption also occurs during low temperature collisions of $O_2$ molecules. Such collisions would be frequent for any density of $O_2$ trapped in voids. Therefore, molecular level modeling of the formation and trapping of $O_2$ is needed.

**5. Gas-phase $O_2$ Sources**

Radiolytically produced $O_2$ molecules ejected from the surfaces of Europa and Ganymede primarily return (e.g., Oza et al. 2019) and rattle around in their highly porous regoliths as discussed earlier. At any time, some of these return into the atmosphere thermalized, while others are transiently trapped at defect sites on the grain surfaces (Johnson et al. 2019). Therefore, these irradiated regoliths are permeated with transiently adsorbed $O_2$, as well as other volatile products. This is in addition to the $O_2$ formed and trapped in voids described above. In this way the sources of $O_2$ to the atmosphere consist of newly produced and ejected $O_2$, ambient $O_2$ in thermal equilibrium with the local regolith and the plasma-induced release of $O_2$ from near surface bubbles.

Globally, the radiolytic production of *new* $O_2$ from ice must be in equilibrium with its destruction in the atmosphere and regolith, and, to a much smaller extent, to downward geologic transport (e.g., Chyba 2000; Johnson et al. 2003). The net destruction must also be in rough equilibrium with the loss to space of its companion decomposition product, $H_2$ (e.g., Szalay et al. 2024). Due to accumulation of $O_2$ in the regolith, thermal desorption dominates direct plasma-induced production of atmospheric $O_2$ (Oza et al. 2018; Johnson et al. 2019). As the satellite surface temperatures vary spatially and the plasma irradiation is non-uniform, the atmospheric column densities of thermalized $O_2$ are asymmetric (Oza et al. 2019)

Thermal desorption of $O_2$ into these atmospheres is suggested by the modeling of observations. A binding energy to grain surfaces, $E_b \sim 0.11$ eV, was estimated from the *seasonal* variability of $O_2$ detected by the Cassini INMS instrument as emanating from Dione and Rhea (Teolis and Waite 2016). A very similar value, $E_b$ ~0.14eV, was estimated (Johnson et al. 2019) to account for the size of the dusk/dawn enhancement in the $O_2$ column density at Europa (Roth et al. 2016; Oza et al. 2018). These extracted values are remarkably consistent with the laboratory studies (He et al. 2016; Laufer et al. 2017) and represent adsorption of $O_2$ at surface defect sites, presumably dangling H bonds. The thermal desorption rate can be roughly approximated by:

$$t_d^{-1} \sim p_d \, \nu_b \exp(-E_b/kT) \qquad (5)$$

where $p_d$ is a desorption probability, $\nu_b$ is the effective vibrational frequency (~$10^{12}$/s; He et al. 2016). Using $E_b$~ 0.14eV, an average temperature on Europa of 100K, then $t_d$ ~ ($10^{-5}$ s) if $p_d$ ~1. This an order of magnitude longer than the time spent between collisions with grains and far shorter than the ballistic hop time above the surface, confirming the rapid recycling of atmospheric $O_2$. Therefore, the average atmospheric $O_2$ source rate ~$10^8$/cm$^2$s estimated at Europa (Oza et al. 2018) suggests a rough *transient* adsorption coverage of $O_2$ on grain surfaces directly exposed to the vacuum of ~ $10^3$ $O_2$/cm$^2$.

Ice having a significant density of voids can become fluid-like (Tachibana et al. 2017) so that under stress or temperature gradients, $O_2$ bubbles can migrate to grain boundaries, as seen in Antarctic ice cores (Miller 1969). Therefore, at an *external* surface, trapped gas can be released aided by sputtering and thermal erosion (Dadic et al. 2010; 2019). Since these processes are slow in large grains, it has been suggested that satellite surface ice with relatively stable inclusions of oxidized products could be delivered geologically to their underground oceans (e.g. Pappalardo et al. 1998) possibly supporting biology (Chyba 2000; Johnson et al. 2003; Hesse et al. 2022).

Although bubble migration to an external surface can contribute to the local atmosphere, destruction by incident plasma primarily determines the steady-state density in ice grains (JJ). That destruction *is* occurring within the photon sampling depth is indicated by the temporal variability observed in the dimer band depths (Spencer et al. 2019; Trumbo et al. 2021). That is, a flux,Φ, of energetic plasma particles can penetrate bubbles of radius $r_b$, in a time, $t_b$~($\pi$ $r_b^2$ Φ )$^{-1}$. This can destroy the trapped $O_2$, cause reactions, as suggested by the presence of $O_3$, and drive products into the ice matrix. Since Φ is roughly proportional to dE/dt in Eq. 1, the flux of plasma particles can destroy bubbles, as well as contributing to their formation (JJ; Appendix A2). Using $r_b$ < ~ 4nm and a flux of highly penetrating plasma onto Europa of ~$10^8$/(cm$^2$s) (Cooper et al. 2001) the destruction time is short compared to its orbital period ($t_b$ <~ 0.1 $t_{orbit}$). Since the plasma flux is variable (Bagenal et al. 2015), at a minimum due to the tilt of Jupiter's dipole (e.g., Milby et al. 2024), and the lifetime of bubbles within the photon sampling depth is short compared to the orbital period, the apparent higher variability in the 'condensed' $O_2$ absorption band depth at Europa as compared to that at Ganymede we suggest is due to the higher radiation rate and lower annealing rate. Indeed, the thermal annealing timescales in Ganymede's equatorial region, following Eqn. 5 for $\nu_b$= $10^{13}$s, $E_a$ = 0.46 eV (Schmitt et al. 2009; Cao et al. 2021 results in a bulk crystallization timescale of only ~ 10-40 minutes or 1-4 × $10^{-3}$ of an orbital period (Table 2, Figure 2), far shorter than Europa's annealing timescales. While recent 3.1 μm JWST observations of Europa's *Tara Regio* and *Powys Regio* (Cartwright et al. 2025) imply recent surface modification of <15 days, our estimates suggest Europa's leading hemisphere equator crystallizes as fast as ~5 days (1.4 Europa orbits). Although far slower than Europa and Ganymede's trailing hemispheres, the leading hemisphere equator is far faster than Northern latitudes at ~100 K approaching ~ 1 Myr timescales, consistent with amorphous ice features in the IR (Cartwright et al. 2025). The thermal region depicted in Figure 2 towards long ~decadal timescales, is consistent with ~5 K surface temperature changes due to Jupiter's eccentric ~11.86 y orbit driving *seasonal* annealing observed on Europa by JWST (Mergny et al. 2025) and Ganymede.

**6. Discussion**

We briefly reviewed aspects of the physics associated with the radiolytically produced $O_2$ observed in the icy regoliths on Europa and Ganymede, and the sources of oxygen into their thin atmospheres. In this, we were guided by modeling of radiation produced voids and bubbles in refractory solids, as well as laboratory data on the radiolysis of ice. Although the latter has been critical to our understanding of spacecraft and telescopic observations, we point out that care must be taken in directly applying this data. That is, as discussed above, the satellite surface ice tends to be highly porous with relatively large grains that experience annealing and relatively low radiation rates, whereas the laboratory samples are typically thin, small grain, microporous samples experiencing relatively high irradiation rates in which annealing due to the background temperature is often irrelevant. Therefore, the direct observation of the radiation-induced dimer bands in laboratory ice samples has been problematic leading to considerable speculation. This is unfortunate as some estimates of the radiolytically produced $O_2$ reservoirs at the icy satellites suggest the 'condensed' $O_2$ is the dominant reservoir of $O_2$ (Johnson et al. 2003). This is the case although bands other than those in the visible have not been identified (Calvin et al. 1996; Calvin 2025), suggesting more detailed modeling of the reflectance spectra is needed.

Observations suggest to us that $O_2$ is trapped in the satellite surfaces in two forms: $O_2$ bubbles ('condensed' $O_2$) and transiently adsorbed $O_2$ on grain surfaces, as indicated by modeling of the dusk/dawn asymmetry (Johnson et al. 2019). Here we also suggest that the observed variability of the 'condensed' $O_2$ bands is due to the destruction of bubbles within the photon sampling depth by the penetrating plasma particles. Therefore, the higher plasma fluxes and lower temperatures on Europa appear consistent with more rapid recycling of smaller bubbles (e.g., Appendix A2) which also contribute to light scattering in the visible. This is opposite to the case at low latitudes on Ganymede which experiences a low flux of primarily penetrating ions and electrons, so that the energy deposition rate per unit path length in the ice is much lower than at Europa, while thermal annealing is more rapid. This is generalized and depicted in Figure 3, starting clockwise from a neutral ice grain (Costello and Oza 2026).

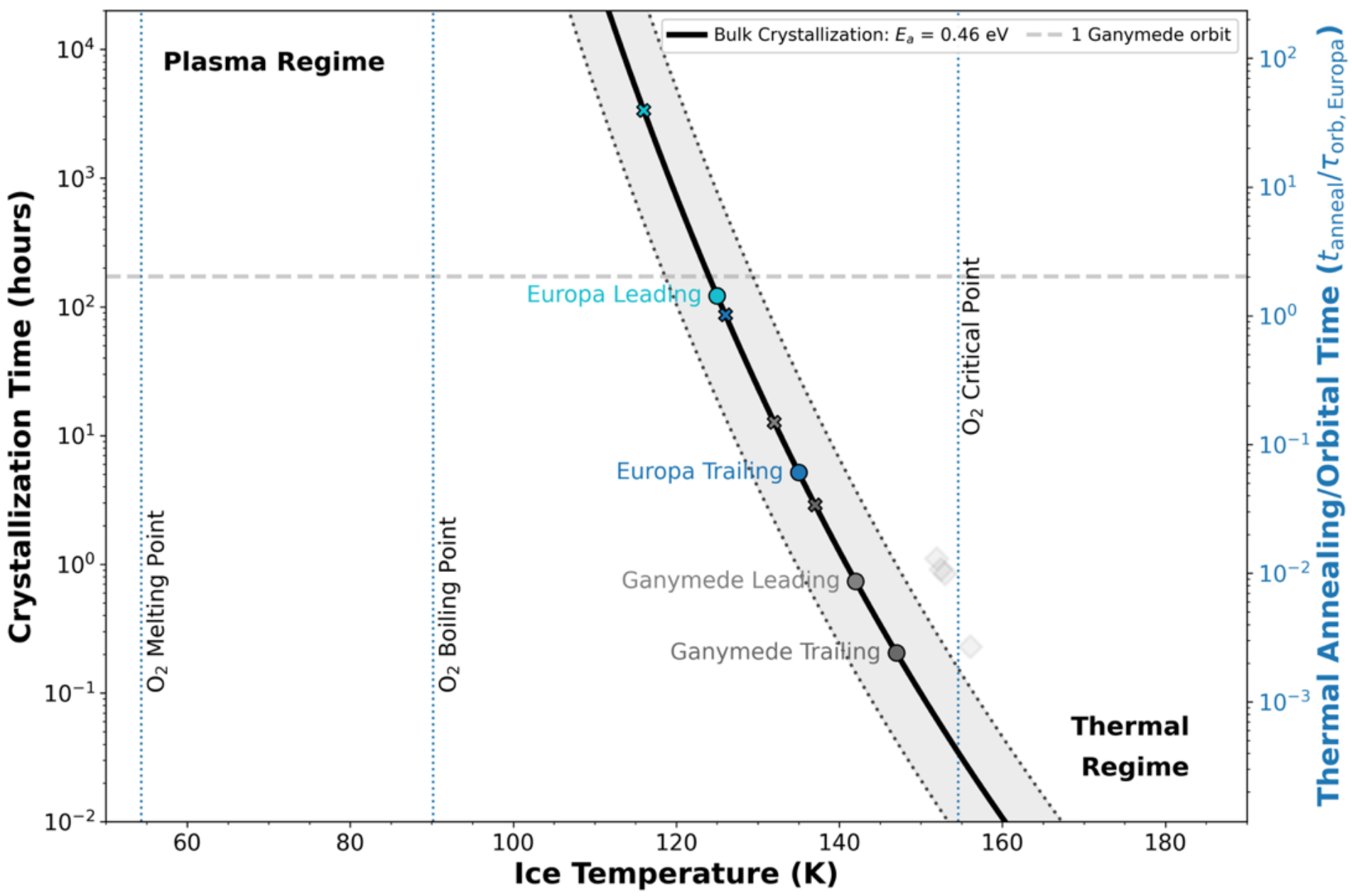

**FIG. 2.** Thermal annealing timescale estimated in units of hours (left) and Europa orbital time (right: vertical axis) following $t_{anneal} \sim 10^{13}$s $e^{(-E/kT))}$ for bulk ice crystallization at $E_a = 0.46$ eV (Schmitt et al. 1989) . The gray region bounds the approximate uncertainty in water ice annealing from experiments (Cao et al. 2021) roughly +/- 0.2 eV (dotted lines). Horizontal lines highlight approximate plasma destruction times for each moon. Europa and Ganymede's equatorial sunlit leading and trailing hemisphere temperatures are indicated for each moon, corresponding to data in Table 2. Two approximate regimes are highlighted: 1. Plasma regime (top left): similar to Europa's northern latitudes where amorphous ice is observed and 2. Thermal regime (lower right): where the annealing time is >1000x orbital and plasma timescales similar to Ganymede's trailing hemisphere. $O_2$ melting, boiling, and critical points are indicated.

Although bubble destruction and migration to grain surfaces can contribute to the atmospheric source rate, the time scales suggested by the variability in the dimer band depth (Spencer et al. 2019) are much longer than the times for the thermal desorption of the returning atmospheric $O_2$. That is, the modeling (Oza et al. 2019) of the observed dusk/ dawn enhancement of the $O_2$–driven aurora at Europa (Roth et al. 2016), which is roughly consistent with plasma observations (Addison et al. 2022), indicates that the dominant supply of atmospheric $O_2$ is thermal desorption. These observations have led to the extraction of adsorption energies on ice grains in their regoliths of ~0.11-0.14 eV (Teolis and Waite, 2016; Johnson et al. 2019), values that we pointed out are roughly consistent with laboratory data (He et al. 2016; Laufer et al. 2017). Finally, we suggested that the previously unexplained plasma-induced $O_2$ 'activation energy', seen extensively in laboratory studies (e.g., Reimann et al. 1984; Orlando and Sieger 2003; Teolis et al. 2017) is determined by the surface tension of ice. This is consistent with trapping of $O_2$ seen at the surface in laboratory samples (Teolis et al. 2009) and the role of interior surfaces in the formation of trapped volatiles. Such aspects must be confirmed by detailed molecular level modeling in preparation for the observations expected from the *Europa Clipper* and *JUICE* missions. Of particular interest is the Moons and Jupiter Imaging Spectrometer (MAJIS) on *JUICE* which may be able to probe oxidant features in shadow (Poulet et al. 2024). Future space observatories such as *Habitable World Observatory* might consider a spectrometer capable of studying the molecular oxygen feature described here with a large-aperture telescope. Thus far space observatories such as *James Webb Space Telescope* (*JWST*) are limited to more complex oxidants on putative ocean worlds: Ariel at the Uranian system (Cartwright et al. 2024a), Europa (Trumbo et al. 2023; Villanueva et al. 2023), Ganymede (Bockelée-Morvan et al. 2024), and Callisto (Cartwright et al. 2024b) probed *in-situ* by *Galileo/NIMS* at Europa (Hansen and McCord 2008) and *Cassini*/*VIMS* at Enceladus (Brown et al. 2006; Combe et al. 2019) as well as $SO_2$ and $SiO_2$ at exoplanetary systems (Inglis et al. 2024; Oza et al. 2026). Understanding the processes discussed above can also help interpret *JWST* observations, suggestive of an active source at Europa (Kadoya et al. 2025) and beyond.

### 7. Summary

Following the early experiments of Brown's and Lanzerotti's group (e.g, Reimann et al. 1984) it became clear that the Jovian magnetospheric plasma ejected $H_2$ and $O_2$ from the surfaces of the icy Galilean satellites. However, the subsequent observations of the so-called 'condensed' $O_2$ bands seen in reflectance spectra at Europa, Ganymede, and Callisto (e.g., Spencer and Calvin 2002) remains an enigma. This is the case because the differences in depths and variability of the bands seen on Europa and Ganymede are *not* simply explained using the available laboratory data.

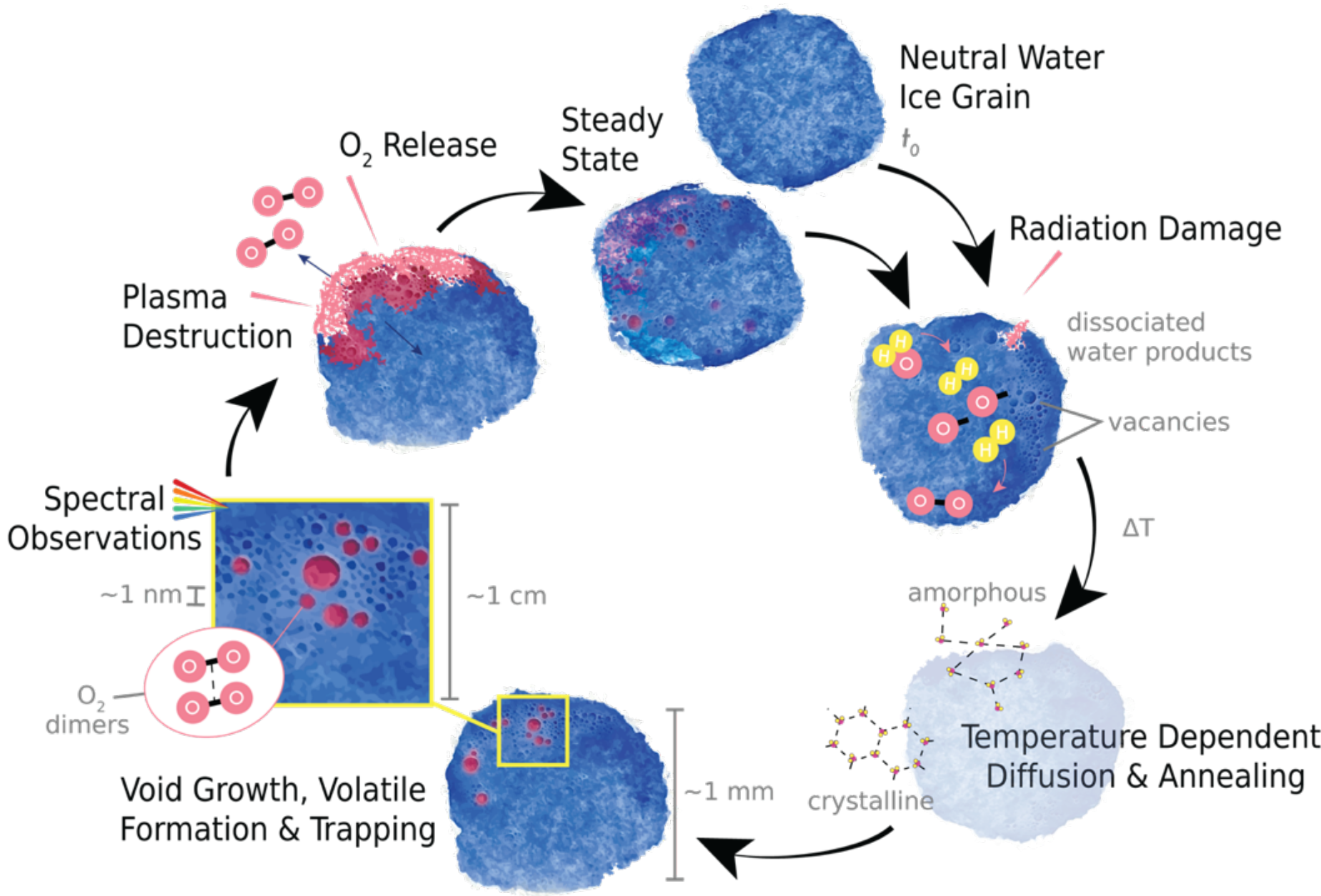

**FIG. 3.** Radiolytic processes in surface ice grains in time. Diurnal temperature variations on Europa (85-130K) and Ganymede (70 – 150 K) and variable energetic particles (~keV→ MeV) lead to observed $O_2$ bubble formation (radiation damage vs annealing ) observed via the 5773 Å $O_2$ dimer absorption bands.

Watercolor painting: Emily Costello (Hawaii Institute of Geophysics and Planetology). N.B. bubbles painted in the figure are natural nitrogen/oxygen bubbles in watercolor paint.

Here we used our understanding of the radiation-induced processes that form, trap and desorb $O_2$ at Europa and Ganymede expanding on earlier work (e.g., JJ). That is, incident plasma particles not only lead to the production of voids in which volatiles can be formed and trapped but *can also lead to their destruction* and to the production of scattering centers that affect the reflected photon sampling depths. Here we described how these processes are affected by thermal annealing rates in competition with the rate of plasma-induced damage production. Although we can understand the differences in the 'condensed' $O_2$ observations on these neighboring icy bodies, a quantitative description of their formation in the ice grains is needed beyond the simple model in Appendix A2 and qualitative depiction in Figure 3 (Costello and Oza 2026). This will require modeling of the molecular physics as well as experiments involving spectroscopy on thicker ice samples exposed to low rates of penetrating electrons or light ions. A quantitative description is needed as other volatile products are also observed to be trapped in near surface ice grains. More importantly, such an understanding could help validate the suggestion that the multiple $O_2$ formed and trapped in the surficial ice might be a potential source of oxygen delivered geologically to their subsurface oceans providing a metabolic source for potential biologic activity underneath the ice as well as the suggestion of ongoing activity at Europa (Kadoya et al. 2025).

**Acknowledgements**

We thank Shane Carberry-Mogan, Sydney A. Willis and the two referees for helpful comments. AVO also thanks Rosaly M.C Lopes for discussions on the importance of oxidants in habitable ocean worlds. AVO expresses gratitude for original artwork by artist and planetary scientist Emily Costello. This work was partly supported by the NAI project "Habitability of Hydrocarbon Worlds: Titan and Beyond" (PI R.M. Lopes).

**Author Disclosure Statement**

No competing financial interests exist.

**Funding Information**

No funding was received for this article.

**References**

Abellan, P., Gautron, E., LaVerne, J.A., 2023. Radiolysis of Thin Water Ice in Electron Microscopy. The Journal of Physical Chemistry C 127, 15336–15345. https://doi.org/10.1021/acs.jpcc.3c02936

Addison, P., Liuzzo, L., Simon, S., 2022. Effect of the Magnetospheric Plasma Interaction and Solar Illumination on Ion Sputtering of Europa's Surface Ice. Journal of Geophysical Research: Space Physics 127, e2021JA030136. https://doi.org/10.1029/2021JA030136

Bagenal, F., Sidrow, E., Wilson, R.J., Cassidy, T.A., Dols, V., Crary, F.J., Steffl, A.J., Delamere, P.A., Kurth, W.S., Paterson, W.R., 2015. Plasma conditions at Europa's orbit. Icarus 261, 1–13. https://doi.org/10.1016/j.icarus.2015.07.036

Bahr, D.A., Famá, M., Vidal, R.A., Baragiola, R.A., 2001. Radiolysis of water ice in the outer solar system: Sputtering and trapping of radiation products. Journal of Geophysical Research: Planets 106, 33285–33290. https://doi.org/10.1029/2000JE001324

Baragiola, R., Bahr, D., 1998. Laboratory studies of the optical properties and stability of oxygen on Ganymede. Journal of Geophysical Research-Planets 103, 25865–25872.

Baragiola, R.A., 2003. Water ice on outer solar system surfaces: Basic properties and radiation effects. Planetary and Space Science 51, 953–961. https://doi.org/10.1016/j.pss.2003.05.007

Behrisch, R., Eckstein, W., 2007. Sputtering by Particle Bombardment, Topics in Applied Physics. Springer, Berlin. https://doi.org/10.1007/978-3-540-44502-9

Bénit, J., Brown, W.L., 1990. Electronic sputtering of oxygen and water molecules prom thin films of water ice bombarded by MeV Ne+ ions. Nuclear Instruments and Methods in Physics Research Section B: Beam Interactions with Materials and Atoms 46, 448–451. https://doi.org/10.1016/0168-583X(90)90745-G

Bockelée-Morvan, D., Poch, O., Leblanc, F., Zakharov, V., Lellouch, E., Quirico, E., de Pater, I., Fouchet, T., Rodriguez-Ovalle, P., Roth, L., Merlin, F., Duling, S., Saur, J., Masson, A., Fry, P., Trumbo, S., Brown, M., Cartwright, R., Cazaux, S., de Kleer, K., Fletcher, L., Milby, Z., Moingeon, A., Mura, A., Orton, G., Schmitt, B., Tosi, F., Wong, M., 2024. A patchy CO2 exosphere on Ganymede revealed by the James Webb Space Telescope. Astronomy & Astrophysics 690. https://doi.org/10.1051/0004-6361/202451599

Brown, R. H., Clark, R. N., Buratti, B. J., Cruikshank, D. P., Barnes, J. W., Mastrapa, R. M. E., Bauer, J., Newman, S., Momary, T., Baines, K. H., Bellucci, G., Capaccioni, F., Cerroni, P., Combes, M., Coradini, A., Drossart, P., Formisano, V., Jaumann, R., Langevin, Y., Matson, D. L., McCord, T. B., Nelson, R. M., Nicholson, P. D., Sicardy, B., & Sotin, C. (2006). Composition and Physical Properties of Enceladus' Surface. Science, 311(5766), 1425–1428.
https://doi.org/10.1126/science.1121031

Brown, W.L., Augustyniak, W.M., Simmons, E., Marcantonio, K.J., Lanzerotti, L.J., Johnson, R.E., Boring, J.W., Reimann, C.T., Foti, G., Pirronello, V., 1982. Erosion and molecule formation in condensed gas films by electronic energy loss of fast ions. Nuclear Instruments and Methods in Physics Research 198, 1–8. https://doi.org/10.1016/0167-5087(82)90043-6

Calvin, W. M., Absorption Coefficients of Condensed O2: Application to Ganymede. Planetary Science Journal, 6, 216, 2025. https://doi.org/10.3847/PSJ/adfc63

Calvin, W.M., Johnson, R.E., Spencer, J.R., 1996. O2 on Ganymede: Spectral characteristics and plasma formation mechanisms. Geophysical Research Letters 23, 673–676. https://doi.org/10.1029/96GL00450

Cao, H.S., 2021. Formation and crystallization of low-density amorphous ice. Journal of Physics D: Applied Physics 54, 203002. https://doi.org/10.1088/1361-6463/abe330

Carberry Mogan, S.R., Tucker, O.J., Johnson, R.E., Roth, L., Alday, J., Vorburger, A., Wurz, P., Galli, A., Smith, H.T., Marchand, B., Oza, A.V., 2022. Callisto's Atmosphere: First Evidence for H2 and Constraints on H2O. Journal of Geophysical Research: Planets 127, e2022JE007294. https://doi.org/10.1029/2022JE007294

Carlson, R.W., Calvin, W.M., Dalton, J.B., Hansen, G.B., Hudson, R.L., Johnson, R.E., McCord, T.B., Moore, M.H., Khurana, K.K., Pappalardo, R.T., McKinnon, W.B., 2017. Europa's Surface Composition. University of Arizona Press, p. 283. https://doi.org/10.2307/j.ctt1xp3wdw.18

Cartwright, R. J., Holler, B. J., Villanueva, G. L., Camarca, M., Faggi, S., Neveu, M., Roth, L., Raut, U., Glein, C. R., Castillo-Rogez, J. C., Malaska, M. J., Bockelée-Morvan, D., Nordheim, T. A., Hand, K. P., Strazzulla, G., Pendleton, Y. J., de Kleer, K., Beddingfield, C. B., de Pater, I., Cruikshank, D. P., & Protopapa, S. (2024). JWST Reveals CO Ice, Concentrated $CO_2$ Deposits, and Evidence for Carbonates Potentially Sourced from Ariel's Interior. *The Astrophysical Journal Letters*, 970, L29. https://doi.org/10.3847/2041-8213/ad566a

Cartwright, R. J., Villanueva, G. L., Holler, B. J., Camarca, M., Faggi, S., Neveu, M., Roth, L., Raut, U., Glein, C. R., Castillo-Rogez, J. C., Malaska, M. J., Bockelée-Morvan, D., Nordheim, T. A., Hand, K. P., Strazzulla, G., Pendleton, Y. J., de Kleer, K., Beddingfield, C. B., de Pater, I., Cruikshank, D. P., & Protopapa, S. (2024). Revealing Callisto's Carbon-rich Surface and $CO_2$ Atmosphere with JWST. The Planetary Science Journal, 5, 60. https://doi.org/10.3847/PSJ/ad23e6

Cartwright, R. J., Hibbitts, C. A., Holler, B. J., Raut, U., Nordheim, T. A., Neveu, M., Protopapa, S., Glein, C. R., Leonard, E. J., Roth, L., Beddingfield, C. B., & Villanueva, G. L. 2025. JWST Reveals Spectral Tracers of Recent Surface Modification on Europa. *Planetary Science Journal* 6, 125. https://doi.org/10.3847/PSJ/adcab9

Cassidy, T.A., Johnson, R.E., 2005. Monte Carlo model of sputtering and other ejection processes within a regolith. Icarus 176, 499–507. https://doi.org/10.1016/j.icarus.2005.02.013

Christianson, D.A., Garrod, R.T., 2021. Chemical Kinetics Simulations of Ice Chemistry on Porous Versus Non-Porous Dust Grains. Frontiers in Astronomy and Space Sciences Volume 8-2021. https://doi.org/10.3389/fspas.2021.643297

Chyba, C.F., 2000. Energy for microbial life on Europa. Nature 403, 381–382. https://doi.org/10.1038/35000281

Combe, J.-P., McCord, T. B., Matson, D. L., Johnson, T. V., Davies, A. G., Scipioni, F., & Tosi, F. (2019). Nature, distribution and origin of $CO_2$ on Enceladus. Icarus, 317, 491–508. https://doi.org/10.1016/j.icarus.2018.08.007

Cooper, J.F., Johnson, R.E., Mauk, B.H., Garrett, H.B., Gehrels, N., 2001. Energetic Ion and Electron Irradiation of the Icy Galilean Satellites. Icarus 149, 133–159. https://doi.org/10.1006/icar.2000.6498

Cooper, P., Johnson, R., Quickenden, T., 2003. Hydrogen peroxide dimers and the production of O2 in icy satellite surfaces. Icarus 166, 444–446. https://doi.org/10.1016/j.icarus.2003.09.008

Dadic, R., Light, B., Warren, S.G., 2010. Migration of air bubbles in ice under a temperature gradient, with application to "Snowball Earth." Journal of Geophysical Research: Atmospheres 115. https://doi.org/10.1029/2010JD014148

Costello, E., Oza, A. (2026). Radiolytic processes in surface ice grains over time. Zenodo. https://doi.org/10.5281/zenodo.19878688

Dadic, R., Schneebeli, M., Wiese, M., Bertler, N.A.N., Salamatin, A.N., Theile, T.C., Alley, R.B., Lipenkov, V.Ya., 2019. Temperature-Driven Bubble Migration as Proxy for Internal Bubble Pressures and Bubble Trapping Function in Ice Cores. Journal of Geophysical Research: Atmospheres 124, 10264–10282. https://doi.org/10.1029/2019JD030891

Davis, M.R., Meier, R.M., Cooper, J.F., Loeffler, M.J., 2021. The Contribution of Electrons to the Sputter-produced O2 Exosphere on Europa. The Astrophysical Journal Letters 908, L53. https://doi.org/10.3847/2041-8213/abe415

de Kleer, K., Milby, Z., Schmidt, C., Camarca, M., Brown, M.E., 2023. The Optical Aurorae of Europa, Ganymede, and Callisto. The Planetary Science Journal 4, 37. https://doi.org/10.3847/PSJ/acb53c

de Koning, M., 2020. Crystal imperfections in ice Ih. The Journal of Chemical Physics 153. https://doi.org/10.1063/5.0019067

Ebert, R. W., Fuselier, S. A., Allegrini, F., Bagenal, F., Bolton, S. J., Clark, G., Connerney, J. E. P., DiBraccio, G. A., Kurth, W. S., Levin, S., McComas, D. J., Montgomery, J., Romanelli, N.,

Sulaiman, A. H., Szalay, J. R., Valek, P., & Wilson, R. J. 2022. Evidence for magnetic reconnection at Ganymede's upstream magnetopause during the PJ34 Juno flyby. Geophysical Research Letters, 49(20), e2022GL099775. https://doi.org/10.1029/2022GL099775

Eldrup, M., 1976. Vacancy migration and void formation in γ-irradiated ice. The Journal of Chemical Physics 64, 5283–5290. https://doi.org/10.1063/1.432157

Famá, M., Loeffler, M.J., Raut, U., Baragiola, R.A., 2010. Radiation-induced amorphization of crystalline ice. Icarus 207, 314–319. https://doi.org/10.1016/j.icarus.2009.11.001

Gittus, J., 1978. Irradiation Effects in Crystalline Solids.

Gusak, A.M., Lutsenko, G.V., Tu, K.N., 2006. Ostwald ripening with non-equilibrium vacancies. Acta Materialia 54, 785–791. https://doi.org/10.1016/j.actamat.2005.09.035

Hall, D., Strobel, D., Feldman, P., Mcgrath, M., Weaver, H., 1995. Detection of an Oxygen Atmosphere on Jupiter Moon Europa. Nature 373, 677–679.

Hand, K., Chyba, C., Carlson, R., Cooper, J., 2006. Clathrate hydrates of oxidants in the ice shell of Europa. Astrobiology 6, 463–482.

Hansen, G. B., and McCord, T. B. (2008). Widespread $CO_2$ and other non-ice compounds on the anti-Jovian and trailing sides of Europa from Galileo/NIMS observations. *Geophysical Research Letters*, 35(1), L01202. https://doi.org/10.1029/2007GL031748

He, J., Acharyya, K., Vidali, G., 2016. Sticking of molecules on nonporous amorphous water ice. The Astrophysical Journal 823, 56. https://doi.org/10.3847/0004-637X/823/1/56

He, Q., Fang, Z., Shoshanim, O., Brown, S.S., Rudich, Y., 2021. Scattering and absorption cross sections of atmospheric gases in the ultraviolet–visible wavelength range (307–725 nm). Atmos. Chem. Phys. 21, 14927–14940. https://doi.org/10.5194/acp-21-14927-2021

Heide, H.G., Zeitler, E., 1985. The physical behavior of solid water at low temperatures and the embedding of electron microscopical specimens. Ultramicroscopy 16, 151–160. https://doi.org/10.1016/0304-3991(85)90070-1

Hesse, M.A., Jordan, J.S., Vance, S.D., Oza, A.V., 2022. Downward Oxidant Transport Through Europa's Ice Shell by Density-Driven Brine Percolation. Geophysical Research Letters 49, e2021GL095416. https://doi.org/10.1029/2021GL095416

Inglis, J., Batalha, N. E., Lewis, N. K., Zellem, R., & … (other authors) (2024). Quartz Clouds in the Dayside Atmosphere of the Quintessential Hot Jupiter HD 189733 b. *The Astrophysical Journal Letters*, 973, L41. https://doi.org/10.3847/2041-8213/ad725e

Jiang, Y., Xin, Y., Liu, W., Sun, Z., Chen, P., Sun, D., Zhou, M., Liu, X., Yun, D., 2022. Phase-field simulation of radiation-induced bubble evolution in recrystallized U–Mo alloy. Nuclear Engineering and Technology 54, 226–233. https://doi.org/10.1016/j.net.2021.07.034

Jin, M., Garrod, R.T., 2020. Formation of Complex Organic Molecules in Cold Interstellar Environments through Nondiffusive Grain-surface and Ice-mantle Chemistry. The Astrophysical Journal Supplement Series 249, 26. https://doi.org/10.3847/1538-4365/ab9ec8

Johnson, R., Lanzerotti, L., Brown, W., 1982. Planetary Applications of Ion Induced Erosion of Condensed-Gas Frosts. Nuclear Instruments & Methods In Physics Research 198, 147–157. https://doi.org/10.1016/0167-5087(82)90066-7

Johnson, R., Oza, A., Leblanc, F., Schmidt, C., Nordheim, T., Cassidy, T., 2019. The Origin and Fate of O2 in Europa's Ice: An Atmospheric Perspective. Space Science Reviews 215. https://doi.org/10.1007/s11214-019-0582-1

Johnson, R., Quickenden, T., 1997. Photolysis and radiolysis of water ice on outer solar system bodies. Journal of Geophysical Research-Planets 102, 10985–10996.

Johnson, R., Quickenden, T., Cooper, P., McKinley, A., Freeman, C., 2003. The production of oxidants in Europa's surface. Astrobiology 3, 823–850. https://doi.org/10.1089/153110703322736123

Johnson, R.E., 1990. Energetic Charged-Particle Interactions with Atmospheres and Surface, Physics and Chemistry in Space. Springer. https://doi.org/10.1007/978-3-642-48375-2

Johnson, R.E., 1985. Polar frost formation on Ganymede. Icarus 62, 344–347. https://doi.org/10.1016/0019-1035(85)90130-7

Johnson, R.E.,, W.A., 1997. O2/O3 Microatmospheres in the Surface of Ganymede. The Astrophysical Journal 480, L79. https://doi.org/10.1086/310614

Kadoya, S., Sekine, Y., & Kodama, T. 2025. Equatorial $CO_2$ Distribution Suggests Active $CO_2$ Supply on Europa. *Planetary Science Journal* 6, 102. https://doi.org/10.3847/PSJ/adc37f

Ketcham, W.M., Hobbs, P.V., 1969. An experimental determination of the surface energies of ice. The Philosophical Magazine: A Journal of Theoretical Experimental and Applied Physics 19, 1161–1173. https://doi.org/10.1080/14786436908228641

King, O., Fletcher, L.N., 2022. Global Modeling of Ganymede's Surface Composition: Near-IR Mapping From VLT/SPHERE. Journal of Geophysical Research: Planets 127, e2022JE007323. https://doi.org/10.1029/2022JE007323

Lanzerotti, L.J., Brown, W.L., Poate, J.M., Augustyniak, W.M., 1978. On the contribution of water products from Galilean satellites to the Jovian magnetosphere. Geophysical Research Letters 5, 155–158. https://doi.org/10.1029/GL005i002p00155

Lanzerotti, L.J., Maclennan, C.G., Brown, W.L., Johnson, R.E., Barton, L.A., Reimann, C.T., Garrett, J.W., Boring, J.W., 1983. Implications of Voyager data for energetic ion erosion of the icy satellites of Saturn. Journal of Geophysical Research: Space Physics 88, 8765–8770. https://doi.org/10.1029/JA088iA11p08765

Laufer, D., Bar-Nun, A., Greenberg, A., 2017. Trapping mechanism of O2 in water ice as first measured by Rosetta spacecraft. Monthly Notices of the Royal Astronomical Society 469, S818–S823. https://doi.org/10.1093/mnras/stx3359

Li, J., Gudipati, M.S., Mishra, Y.N., Liang, M.-C., Yung, Y.L., 2022. Oxidant generation in the ice under electron irradiation: Simulation and application to Europa. Icarus 373, 114760. https://doi.org/10.1016/j.icarus.2021.114760

Li, J., Gudipati, M.S., Yung, Y.L., 2020. The influence of Europa's plumes on its atmosphere and ionosphere. Icarus 352, 113999. https://doi.org/10.1016/j.icarus.2020.113999

Ligier, N., Paranicas, C., Carter, J., Poulet, F., Calvin, W., Nordheim, T., Snodgras, C., Ferellec, L., 2019. Surface composition and properties of Ganymede: Updates from ground-based observations with the near-infrared imaging spectrometer SINFONI/VLT/ESO. Icarus 333, 496–515. https://doi.org/10.1016/j.icarus.2019.06.013

Liuzzo, L., Poppe, A. R., Paranicas, C., Nénon, Q., Fatemi, S., & Simon, S., 2020. Variability in the Energetic Electron Bombardment of Ganymede. Journal of Geophysical Research: Space Physics, 125(9), e2020JA028347. https://doi.org/10.1029/2020JA028347

Mamo, B. D., Raut, U., Teolis, B., Erwin, T. P., Cartwright, R. J., Protopapa, S., Retherford, K. D., & Nordheim, T. A. 2025. La-boratory Investigation of $CO_2$-Driven Enhancement of Radiolytic $H_2O_2$ on Europa and Other Icy Moons. *Planetary Science Journal* 6, 170. https://doi.org/10.3847/PSJ/ade3d8

Matich, A., Bakker, M., Lennon, D., Quickenden, T., Freeman, C., 1993. O2 luminescence from UV-excited H2O and D2O ices. Journal of Physical Chemistry 97, 10539–10553. https://doi.org/10.1021/j100143a007

McKellar, A.R.W., Rich, N.H., Welsh, H.L., 1972. Collision-Induced Vibrational and Electronic Spectra of Gaseous Oxygen at Low Temperatures. Canadian Journal of Physics 50, 1–9. https://doi.org/10.1139/p72-001

Mergny, C., Schmidt, F., & Keil, F. 2025. The blinking crystallinity of Europa: A competition between irradiation and thermal alteration. *Icarus* 441, 116700. https://doi.org/10.1016/j.icarus.2025.116700

Michalsky, J., Beauharnois, M., Berndt, J., Harrison, L., Kiedron, P., Min, Q., 1999. O2-O2 absorption band identification based on optical depth spectra of the visible and near-infrared. Geophysical Research Letters 26, 1581–1584. https://doi.org/10.1029/1999GL900267

Mifsud, D.V., Hailey, P.A., Herczku, P., Juhász, Z., Kovács, S.T.S., Sulik, B., Ioppolo, S., Kaňuchová, Z., McCullough, R.W., Paripás, B., Mason, N.J., 2022. Laboratory experiments on the radiation astrochemistry of water ice phases. The European Physical Journal D 76, 87. https://doi.org/10.1140/epjd/s10053-022-00416-4

Migliorini, A., Kanuchova, Z., Ioppolo, S., Barbieri, M., Jones, N.C., Hoffmann, S.V., Strazzulla, G., Tosi, F., Piccioni, G., 2022. On the origin of molecular oxygen on the surface of Ganymede. Icarus 383, 115074. https://doi.org/10.1016/j.icarus.2022.115074

Milby, Z., Katherine, de K., Schmidt, C., Leblanc, F., 2024. Short-timescale Spatial Variability of Ganymede's Optical Aurora. The Planetary Science Journal 5, 153. https://doi.org/10.3847/PSJ/ad49a2

Miller, S.L., 1969. Clathrate Hydrates of Air in Antarctic Ice. Science 165, 489–490. https://doi.org/10.1126/science.165.3892.489

Noll, K., Johnson, R., Lane, A., Domingue, D., Weaver, H., 1996. Detection of ozone on Ganymede. Science 273, 341–343. https://doi.org/10.1126/science.273.5273.341

Nordheim, T., Hand, K., Paranicas, C., 2018. Preservation of potential biosignatures in the shallow subsurface of Europa. Nature Astronomy 2, 673–679. https://doi.org/10.1038/s41550-018-0499-8

Okada, N., Ohkubo, T., Maruyama, I., Murakami, K., Suzuki, K., 2020. Characterization of irradiation-induced novel voids in α-quartz. AIP Advances 10. https://doi.org/10.1063/5.0029299

Orlando, T., Sieger, M., 2003. The role of electron-stimulated production of O2 from water ice in the radiation processing of outer solar system surfaces. Surface Science 528, 1–7. https://doi.org/10.1016/S0039-6028(02)02602-X

Oza, A.V., Johnson, R., 2024. Common origin of trapped volatiles in oxidized icy moons and comets. Icarus 411. https://doi.org/10.1016/j.icarus.2024.115944

Oza, A.V, Johnson, R., Leblanc, F., 2018. Dusk/dawn atmospheric asymmetries on tidally-locked satellites: O2 at Europa. Icarus 305, 50–55. https://doi.org/10.1016/j.icarus.2017.12.032

Oza, A.V, Leblanc, F., Johnson, R., Schmidt, C., Leclercq, L., Cassidy, T., Chaufray, J., 2019. Dusk over dawn O2 asymmetry in Europa's near-surface atmosphere. Planetary and Space Science 167, 23–32. https://doi.org/10.1016/j.pss.2019.01.006

Oza, A. V., Gebek, A., Meyer zu Westram, M., Tokadjian, A., Piro, A. L., Hu, R., Unni, A., Chari, R., Bello-Arufe, A., Schmidt, C. A., Louca, A. J., Miguel, Y., Estrela, R., Yang, J., Damiano, M., Hasegawa, Y., Welbanks, L., Powell, D., Garg, R., Gupta, P., Yung, Y. L., & Lopes, R. M. C. (2026). Volcanic Satellites Tidally Venting Na, K, $SO_2$ in Optical & Infrared Light. *Monthly Notices of the Royal Astronomical Society*. (in press) https://doi:10.1093/mnras/staf1526

Pappalardo, R.T., Head, J.W., Greeley, R., Sullivan, R.J., Pilcher, C., Schubert, G., Moore, W.B., Carr, M.H., Moore, J.M., Belton, M.J.S., Goldsby, D.L., 1998. Geological evidence for solid-state convection in Europa's ice shell. Nature 391, 365–368. https://doi.org/10.1038/34862

Patterson, J.D., Saltzman, E.S., 2021. Diffusivity and Solubility of H2 in Ice Ih: Implications for the Behavior of H2 in Polar Ice. Journal of Geophysical Research: Atmospheres 126, e2020JD033840. https://doi.org/10.1029/2020JD033840

Petrenko, V.F., Whitworth, R.W., 2010. Physics of Ice. Oxford University Press. https://doi.org/10.1093/acprof:oso/9780198518945.001.0001

Poulet, F., Piccioni, G., Langevin, Y., Dumesnil, C., Tommasi, L., Carlier, V., Filacchione, G., Amoroso, M., Arondel, A., D'Aversa, E., Barbis, A., Bini, A., Bolsee, D., Bousquet, P., Caprini, C., Carter, J., Dubois, J., Condamin, M., Couturier, S., Dassas, K., Dexet, M., Fletcher, L., Grassi, D., Guerri, I., Haffoud, P., Larigauderie, C., Le Du, M., Mugnuolo, R., Pilato, G., Rossi, M., Stefani, S., Tosi, F., Vincendon, M., Zambelli, M., Arnold, G., Bibring, J., Biondi, D., Boccaccini, A., Brunetto, R., Carapelle, A., González, M., Hannou, C., Karatekin, O., Le Cle'ch, J., Leyrat, C., Migliorini, A., Nathues, A., Rodriguez, S., Saggin, B., Sanchez-Lavega, A., Schmitt, B., Seignovert, B., Sordini, R., Stephan, K., Tobie, G., Zambon, F., Adriani, A., Altieri, F., Bockelée, D., Capaccioni, F., De Angelis, S., De Sanctis, M., Drossart, P., Fouchet, T., Gerard, J., Grodent, D., Ignatiev, N., Irwin, P., Ligier, N., Manaud, N., Mangold, N., Mura, A., Pilorget, C., Quirico, E., Renotte, E., Strazzulla, G., Turrini, D., Vandaele, A., Carli, C., Ciarniello, M., Guerlet, S., Lellouch, E., Mancarella, F., Morbidelli, A., Le Mouélic, S., Raponi, A., Sindoni, G., Snels, M., 2024. Moons and Jupiter Imaging Spectrometer (MAJIS) on Jupiter Icy Moons Explorer (JUICE). Space Science Reviews 220. https://doi.org/10.1007/s11214-024-01057-2

Ramachandran, R., Meka, J.K., Rahul, K.K., Khan, W., Lo, J.I., Cheng, B.M., Mifsud, D.V., Rajasekhar, B.N., Das, A., Hill, H., Janardhan, P., Bhardwaj, A., Mason, N.J., Sivaraman, B., 2024. Ultraviolet spectrum reveals the presence of ozone on Jupiter's moon Callisto. Icarus 410, 115896. https://doi.org/10.1016/j.icarus.2023.115896

Reimann, C.T., Boring, J.W., Johnson, R.E., Garrett, J.W., Farmer, K.R., Brown, W.L., Marcantonio, K.J., Augustyniak, W.M., 1984. Ion-induced molecular ejection from D2O ice. Surface Science 147, 227–240. https://doi.org/10.1016/0039-6028(84)90177-8

Roth, L., Saur, J., Retherford, K., Strobel, D., Feldman, P., McGrath, M., Spencer, J., Blöcker, A., Ivchenko, N., 2016. Europa's far ultraviolet oxygen aurora from a comprehensive set of HST observations. Journal of Geophysical Research-Space Physics 121, 2143–2170. https://doi.org/10.1002/2015JA022073

Schaible, M., Johnson, R., Zhigilei, L., Piqueux, S., 2017. High energy electron sintering of icy regoliths: Formation of the PacMan thermal anomalies on the icy Saturnian moons. Icarus 285, 211–223. https://doi.org/10.1016/j.icarus.2016.08.033

Schmitt, B., Espinasse, S., Grim, R. J. A., Greenberg, J. M., & Klinger, J. 1989. Laboratory studies of cometary ice analogues. Proceedings of an International Workshop on Physics and Mechanics of Cometary Materials, eds. J. J. Hunt & T. D. Guyenne, ESA SP-302, Noordwijk: European Space Agency, pp. 65–69

Sieger, M., Simpson, W., Orlando, T., 1998. Production of O2 on icy satellites by electronic excitation of low-temperature water ice. Nature 394, 554–556.

Spencer, J.R., 1987. Icy Galilean satellite reflectance spectra: Less ice on Ganymede and Callisto? Icarus 70, 99–110. https://doi.org/10.1016/0019-1035(87)90077-7

Spencer, J.R., Calvin, W.M., 2002. Condensed O-2 on Europa and Callisto. Astronomical Journal 124, 3400–3403. https://doi.org/10.1086/344307

Spencer, J.R., Calvin, W.M., Person, M.J., 1995. Charge-Coupled-Device Spectra of the Galilean Satellites - Molecular-Oxygen on Ganymede. Journal of Geophysical Research-Planets 100, 19049–19056. https://doi.org/10.1029/95JE01503

Spencer, J.R., W.M. Grundy, and C. Schmidt 2019. Rapid Temporal Variability of Condensed Oxygen on Europa? EPSC-DPS, Geneva, Switzerland, id. EPSC-DPS2019-935-1

Stephan, K., Hibbitts, C., Jaumann, R., 2020. H2O-ice particle size variations across Ganymede's and Callisto's surface. Icarus 337. https://doi.org/10.1016/j.icarus.2019.113440

Szalay, J.R., Allegrini, F., Ebert, R.W., Bagenal, F., Bolton, S.J., Fatemi, S., McComas, D.J., Pontoni, A., Saur, J., Smith, H.T., Strobel, D.F., Vance, S.D., Vorburger, A., Wilson, R.J., 2024. Oxygen production from dissociation of Europa's water-ice surface. Nature Astronomy 8, 567–576. https://doi.org/10.1038/s41550-024-02206-x

Tachibana, S., Kouchi, A., Hama, T., Oba, Y., Piani, L., Sugawara, I., Endo, Y., Hidaka, H., Kimura, Y., Murata, K., Yurimoto, H., Watanabe, N., 2017. Liquid-like behavior of UV-irradiated interstellar ice analog at low temperatures. Science Advances 3. https://doi.org/10.1126/sciadv.aao2538

Teolis, B., Plainaki, C., Cassidy, T., Raut, U., 2017. Water Ice Radiolytic O2, H2, and H2O2 Yields for Any Projectile Species, Energy, or Temperature: A Model for Icy Astrophysical Bodies. Journal of Geophysical Research-Planets 122, 1996–2012. https://doi.org/10.1002/2017JE005285

Teolis, B., Shi, J., Baragiola, R., 2009. Formation, trapping, and ejection of radiolytic O2 from ion-irradiated water ice studied by sputter depth profiling. Journal of Chemical Physics 130. https://doi.org/10.1063/1.3091998

Teolis, B., Waite, J., 2016. Dione and Rhea seasonal exospheres revealed by Cassini CAPS and INMS. Icarus 272, 277–289. https://doi.org/10.1016/j.icarus.2016.02.031

Tinner, C., Galli, A., Baer, F., Pommerol, A., Rubin, M., Vorburger, A., Wurz, P., 2024. Electron-Induced Radiolysis of Water Ice and the Buildup of Oxygen. Journal of Geophysical Research-Planets 129. https://doi.org/10.1029/2024JE008393

Trumbo, S., Brown, M., Bockelée-Morvan, D., de Pater, I., Fouchet, T., Wong, M., Cazaux, S., Fletcher, L., de Kleer, K., Lellouch, E., Mura, A., Poch, O., Quirico, E., Rodriguez-Ovalle, P., Showalter, M., Tiscareno, M., Tosi, F., 2023. Hydrogen peroxide at the poles of Ganymede. Science Advances 9. https://doi.org/10.1126/sciadv.adg3724

Trumbo, S., Brown, M., Hand, K., 2019. H2O2 within Chaos Terrain on Europa's Leading Hemisphere. Astronomical Journal 158. https://doi.org/10.3847/1538-3881/ab380c

Trumbo, S.K., Brown, M.E., Adams, D., 2021. The Geographic Distribution of Dense-phase O2 on Ganymede. The Planetary Science Journal 2, 139. https://doi.org/10.3847/PSJ/ac0cee

Vidal, R., Bahr, D., Baragiola, R., Peters, M., 1997. Oxygen on Ganymede: Laboratory studies. Science 276, 1839–1842. https://doi.org/10.1126/science.276.5320.1839

Villanueva, G., Hammel, H., Milam, S., Faggi, S., Kofman, V., Roth, L., Hand, K., Paganini, L., Stansberry, J., Spencer, J., Protopapa, S., Strazzulla, G., Cruz-Mermy, G., Glein, C., Cartwright, R., Liuzzi, G., 2023. Endogenous $CO_2$ ice mixture on the surface of Europa and no detection of plume activity. Science 381, 1305-+. https://doi.org/10.1126/science.adg4270

Vook, F., Birnbaum, H., Blewitt, T., Brown, W., Corbett, J., Crawford, J., Goland, A., Kulcinski, G., Robinson, M., Seidman, D., Young, F., 1975. Report to American-Physical-Society by Study-Group on Physics Problems Relating to Energy Technologies - Radiation Effects on Materials. Reviews of Modern Physics 47, S1–S44. https://doi.org/10.1103/RevModPhys.47.S1.3

Waite, J.J., Greathouse, T., Mogan, S., Sulaiman, A., Valek, P., Allegrini, F., Ebert, R., Gladstone, G., Kurth, W., Connerney, J., Clark, G., Bagenal, F., Duling, S., Romanelli, N., Bolton, S., Vorburger, A., Paranicas, C., Kollmann, P., Mauk, B., Hansen, C., Buccino, D., Johnson, R., Wilson, R., Teolis, B., 2024. Magnetospheric-Ionospheric-Atmospheric Implications From the Juno Flyby of Ganymede. Journal of Geophysical Research-Planets 129. https://doi.org/10.1029/2023JE007859

Warren, S., 2019. Optical properties of ice and snow. Philosophical Transactions of the Royal Society A-Mathematical Physical and Engineering Sciences 377. https://doi.org/10.1098/rsta.2018.0161

Wiedersich, H., 1972. On the theory of void formation during irradiation. Radiation Effects 12, 111–125. https://doi.org/10.1080/00337577208231128

Wu, P., Trumbo, S.K., Brown, M.E., de Kleer, K., 2024. Europa's H2O2: Temperature Insensitivity and a Correlation with CO2. The Planetary Science Journal 5, 220. https://doi.org/10.3847/PSJ/ad7468

# Appendix

In Appendix A1 we propose the process determining the 'unexplained' activation energy seen in studies of ejection of $O_2$ from irradiated ice, and in (A2) we describe the dependence of the steady state void size on the rate of irradiation of ice grains.

### Appendix A1

The temperature dependence of the $O_2$ ejection yield from irradiated ice is suggestive of an activated process but with an 'unexplained' activation energy (Reimann et al. 1984) (~0.06 eV, depending somewhat on sample preparation). Here we suggest that it could be related to the transient heat pulse produced by an incident particle as thermal conduction varies *inversely* with temperature over the range of interest. However, grain boundaries play a critical role in trapping and release of volatiles (e.g., Jiang et al. 2022). In addition, $O_2$ has been shown to be trapped near irradiated lab sample surfaces before being ejected (Teolis et al. 2006). Here we propose that the surface energy of ice, $\gamma_s$ ~0.07-0.1J/m$^2$ (Ketcham & Hobbs 1969; Petrenko and Whitworth 1999), which is ~0.04-0.07 eV/$H_2O$, likely determines the activation energy' extracted in laboratory studies. Confirmation will require molecular level modeling of radiation damage and transport of defects in ice grains.

### Appendix A2

Using simple rate equations we derive a rough model of the dependence of the size of the average radius, $r_{vo}$, of voids produced in ice on the energy deposition rate, dE/dt, by incident plasma particles. This requires accounting for *both formation and destruction* of voids. A better description would require a detailed chemical kinetic model (e.g., Wiedersich, 1972; Christianson and Garrod 2021) or a phase field model (e.g., Jiang et al. 2022) for ice under irradiation.

Accounting only for the radiation-induced production of vacancies above the thermal background vacancy density, a rate equation for the enhanced vacancy density, $n_v$, under uniform irradiation is

$$dn_v/dt \sim G_v\, n\, dE/dt - k_{v,i}\, n_v\, n_i - k_{v,vo}\, n_v\, n_{vo} \qquad .\text{A1}$$

Here n, $n_i$ and $n_{vo}$ are the density of water molecules, interstitials and voids, with $G_v$ the number of vacancies formed per 100eV deposited (a so-called G-value). In addition, $k_{vi}$ and $k_{v,vo}$ are the vacancy reaction rates with interstitials, leading to annihilation, and with a background of voids, leading to void growth. There are, of course, coupled equations for the interstitials formed by the incident plasma, which we suggest can lead to the formation of $O_2$ on surfaces in voids as well as to other trapped species. Since there are several types of interstitials produced in ice, for simplicity, we focus on vacancies accumulating in voids to obtain a rough estimate of the dependence of void radius on the energy deposition rate.

Assuming that vacancy/interstitial annihilation close to the track produced by an incident particle dominates, and there is a relatively low density of voids, then under steady irradiation, $dn_{vo}/dt \sim 0$, $dn_v/dt \sim 0$ and $n_i \sim n_v$. Under these assumptions Eq. A1 gives an average, steady state vacancy density

$$n_v \sim [G_v\, n\, (dE/dt)/k_{v,i}]^{0.5} \qquad .\text{A2}$$

Assuming the volume of a vacancy is ~1/n, the growth rate of voids from Eq. A1 is $\sim(1/n)(k_{v,vo}\, n_v)$. Also, assuming the void size change due to destruction depends on the probability of an impact of a void by penetrating radiation, as discussed in the text, the impact rate is the void cross sectional area, $A_vo$, times the relevant particle flux, $\Phi$.

Writing the vacancy removal per impact as a G-value, $G_d$, the volume loss rate of voids is $\sim(1/n)\,(G_d\, \Delta E)*(A_v\, \Phi)$ where $\Delta E$ is the energy per incident particle deposited near the void. Writing $\Delta E \sim (n\, S\, r_{vo})$, where S is the stopping power of the penetrating particle and $r_{vo}$ the void radius, results in a volume loss rate of $\sim[A_{vo}G_d\; r_{vo}\, dE/dt]$, where $dE/dt = (S\, \Phi)$. When the destruction and growth rate are roughly equal in steady state, then

$$[A_{vo}G_d\, r_{vo}\, dE/dt] \sim \sim (k_{v,vo}/n)[G_v\, n\, (dE/dt)/k_{v,i}]^{0.5} \qquad .\ \text{A3}$$

Writing the rate constant for vacancy absorption by a void as $k_{v,vo} \sim A_{vo} \times \langle v_v \rangle$, where $\langle v_v \rangle$ is an effective vacancy diffusion speed and rearranging Eq. A3, gives

$$r_{vo} \sim (\langle v_v \rangle/G_d)[G_v\, /(n\, dE/dt\; k_{v,i})]^{0.5} \qquad \text{A4}$$

This expression for the average void radius under steady irradiation has the dependence on dE/dt given in Eq. 1 for the semi-empirical radius of bubbles $r_b$ (i.e., voids containing volatiles). The average, steady state is consistent with a rough *inverse dependence* on the energy deposition rate, but with a larger power than that suggested for gas containing bubbles in Gittus (1978). The dependence on temperature is primarily in the vacancy diffusion velocity with radiation type affecting the G-values. This model is, of course, oversimplified as there is a distribution of void and bubble sizes which requires more detailed modeling and will become important as better data is obtained.